\documentclass[twocolumn,superscriptaddress,showpacs,nofootinbib,pre,showkeys]{revtex4} 

\newcommand{\outpm}[1]{\ensuremath {}^{\pm}#1^{out}}
\newcommand{\inpm}[1]{\ensuremath {}^{\pm}#1^{in}}

\usepackage{graphicx}
\usepackage{hyperref}
\usepackage{amssymb,amsfonts,amsmath, amsthm}

\begin{document}

\title{Community detection in networks with positive and negative links}
\date{21 September 2009}
\author{V.A. Traag}
\email[Corresponding author:~]{vincent.traag@uclouvain.be}
\affiliation{Department of Mathematical Engineering, Universit\'e Catholique de
Louvain, B\^atiment Euler, Avenue G. Lema\^itre 4, B-1348 Louvain-la-neuve, Belgium}
\affiliation{Department of Sociology and Anthropology, Faculty of Social and
Behavioural Sciences, University of Amsterdam, Oudezijds Achterburgwal 185,
1012 DK Amsterdam, the Netherlands}
\author{Jeroen Bruggeman}
\affiliation{Department of Sociology and Anthropology, Faculty of Social and
Behavioural Sciences, University of Amsterdam, Oudezijds Achterburgwal 185,
1012 DK Amsterdam, the Netherlands}
\keywords{complex networks; community detection; modularity; negative links; social balance theory; international relations}
\pacs{89.75.Hc, 89.65.-s}

\begin{abstract}
  Detecting communities in complex networks accurately is a prime challenge,
  preceding further analyses of network characteristics and dynamics. Until
  now, community detection took into account only positively valued links,
  while many actual networks also feature negative links. We extend an existing
  Potts model to incorporate negative links as well, resulting in a method
  similar to the clustering of signed graphs, as dealt with in social balance
  theory, but more general. To illustrate our method, we applied it to a
  network of international alliances and disputes. Using data from 1993--2001,
  it turns out that the world can be divided into six power blocs similar to
  Huntington's civilizations, with some notable exceptions.
\end{abstract}

\maketitle

\section{Introduction}

  Many complex phenomena can be represented as networks, and subsequently be
  analyzed fruitfully~\cite{boccaletti06, dorogovtsev07, bruggeman08}.  One of
  the first targets of network analysis is the detection of communities on the
  basis of the links, i.e.~the possibly valued, or weighted, arcs or edges that
  connect the nodes.  Intuitively, an assignment of nodes to communities should
  be such that links within communities are relatively dense and between
  communities relatively sparse.  This means we should compare actual densities
  to expected densities of links within and between communities.  Furthermore,
  since nodes, for example humans or proteins, can be members of different
  communities at the same time, e.g. organizations or protein complexes,
  respectively, the assignment should allow for the possibility that communities
  overlap.
	
  In approaches to find appropriate community assignments, much progress has
  been made in recent years~\cite{duch, newman_eigenvector, palla05} by using a
  concept known as {\em modularity}~\cite{girvan_newman02}.  While current
  modularity approaches take for granted that links are positively valued,
  representing bonds or attraction, scientists in numerous fields grapple with
  networks that also have \emph{negative} links that represent repel, conflict,
  or opposition, for example in neural networks, semantic webs, genetic
  regulatory networks, and last but certainly not least, in social networks. 
	
  In this paper, we generalize an existing Potts
  model~\cite{reichardt_spinglass} for positive links to incorporate negative
  links as well.  We will follow the intuition that the assignment of nodes
  related by negative links should be done the opposite way of positive links,
  with negative links sparse within and more dense between communities,
  generalizing an old idea from social balance theory~\cite{harary53}.  Finally,
  we apply our approach to a network of conflicts and alliances between
  countries.
	
  Recently, it was shown that modularity might miss small communities embedded
  in larger ones~\cite{fortunato07}, and is less accurate if the actual
  communities are highly different in size~\cite{du08}.  Our method has two
  balancing parameters that address this problem to some
  extent~\cite{kumpala07}. Yet community detection through modularity remains a
  global rather than a local approach.
	
\section{Problem statement}

  We consider a directed graph $G$ with $n$ nodes and $m$ links, which can be
  easily generalized to weighted graphs.  We denote the total number of positive
  links in $G$ as $m^+$ and the number of negative links as $m^{-}$, hence $m =
  m^+ + m^-$.  We define the entries of the adjacency matrix of $G$ as follows: if
  a positive link is present from node $i$ to node $j$, $A_{ij} = 1$, if a
  negative link is present, $A_{ij} = -1$, and $A_{ij} = 0$ otherwise. For a
  weighted graph the link values, or weights, are denoted by $w_{ij}$. We separate
  the negative and positive links by setting $A^{+}_{ij} = A_{ij}$ if $A_{ij} > 0$
  and zero otherwise, and $A^-_{ij} = -A_{ij}$ if $A_{ij} < 0$ and zero otherwise,
  so $A = A^+ - A^-$. The positive and negative in- and outdegrees of $i$ are
  defined as
  \begin{equation}
          \begin{array}{rclcrcl}
                  \outpm{k}_i & = & \sum_{j} A^\pm_{ij} & ~ & \inpm{k}_i & = & \sum_{j} A^\pm_{ji} \\
          \end{array}
  \end{equation}

  Our challenge is to assign each node $i$ to one of $c$ communities $\sigma_i \in
  \{1, \ldots, c\}$. A complete configuration of community assignments is denoted
  by $\{\sigma\}$, which assigns each node $i=1,\ldots,n$ to a community
  $\sigma_1,\ldots,\sigma_n$. 

\section{Social Balance}

  The challenge of community detection in networks with positive and negative
  links was first addressed by social balance theory, which has its origins in
  cognitive dissonance theory~\cite{heider} from the 1950s. This theory is based
  on the notion that if two people are positively related, their attitudes
  toward a third person should match. For example, if Harry and Mary are
  positively related as friends, and both of them are related to John, they
  should both be related to him either positively or negatively. In either case
  their triad is said to be socially balanced. If Harry has a positive
  relationship with John while Mary is negatively related to John or vice versa,
  their triad is socially unbalanced. If all triads in a network are balanced,
  the network is said to be balanced. This definition was later generalized to
  cycles, a triad (a cycle of length 3) being a special case.

  The question whether a balanced network can be divided into separate parts
  arises naturally. The challenge is to define clusters of nodes such that there
  are only positive link within clusters and negative links are between
  clusters.  It was proven~\cite{harary53} that if a connected network is
  balanced, it can be split into two opposing clusters (and vice versa). 

  However, there is an ambiguous case. If a triad has only negative
  relationships, it is neither balanced nor can it be split into two clusters.
  But it can obviously be split into three clusters. In order to accommodate for
  this possibility, the definition of balance was generalized to
  $k$-balance~\cite{davis67,cartwright_harary68}. A network is $k$-balanced if
  it can be divided into $k$ different clusters, each cluster having only
  positive links within itself, and negative links with other clusters. It can
  be proven~\cite{davis67} that a network is balanced if and only if it contains
  no cycles with exactly one negative link. The intuition is simple. Suppose
  there is a cycle $v_1v_2\ldots v_kv_1$ with one negative link, say between
  $v_1$ and $v_k$, and only positive links between the remainder nodes, then $1$
  and $k$ are both positively and negatively connected, and the cycle is
  unbalanced. But if in this cycle there is also a negative link between $i$ and
  $j$, and $1 \leq i < j \leq k$, we can split the cycle in two parts, one
  cluster from $1$ to $i$ and one from $j$ to $k$. If there are more than two
  negative links, we can split up the cycle analogously into more clusters. 

\section{Frustration}

  In reality, however, social networks are rarely, if ever, fully $k$-balanced.
  The question then becomes whether we can still cluster nodes. Obviously, there
  are some links that make a network unbalanced. The number of such links can be
  expressed as an amount of \emph{frustration}. Links that contribute to
  frustration are negative links within clusters and positive links between
  clusters. So the following expression should be minimized
  \begin{equation}
      \sum_{ij} \lambda A^-_{ij}\delta(\sigma_i,\sigma_j) \displaystyle + (1 - \lambda)A^+_{ij}(1 - \delta(\sigma_i,\sigma_j)),
  \end{equation}
  where $\delta(\sigma_i,\sigma_j) = 1$ if $\sigma_i = \sigma_j$ and zero
  otherwise, and  $\lambda$ is a parameter through which the contribution of the
  two types of links can be balanced. This is the approach taken in
  \cite{doreian96,jensen2006}.

  The objective, then, is to find a clustering $\{\sigma\}$ such that the
  frustration is minimized. Since the term $(1 - \lambda)A^+_{ij}$ does not depend
  on the specific configuration and is therefore irrelevant for finding the
  minimum, we can simplify the above expression to
  \begin{equation}
   \sum_{ij} (\lambda A^-_{ij} - (1 - \lambda)A^+_{ij})\delta(\sigma_i,\sigma_j),
  \end{equation}
  We can now see that only for $\lambda=1/2$ we retrieve $A = A^+ - A^-$, up to a
  multiplicative constant of $2$. Using any other value for $\lambda$ would change
  the minimum found, but changing $\lambda$ is the same as altering the (weights
  of the) original network. Setting $\lambda=1/2$ accordingly, we can simplify
  further, and now define frustration as
  \begin{equation}
  F(\{\sigma\}) = - \sum_{ij} A_{ij}\delta(\sigma_i,\sigma_j).
  \label{equ:frustration}
  \end{equation}

  However, frustration does not generalize to a network with only positive links.
  In that case, frustration groups together all nodes into one cluster. Even if
  there are some negative links, frustration will cluster together very sparsely
  connected nodes. It's therefore clear that this approach does not match with
  current methods of community detection. Preferably, there should not be a
  distinction between methods for positive and others for negative links, but
  there should be one method for both.

\section{Modularity}

  In approaches to find appropriate community assignments in networks with only
  positive links, much progress has been made recently~\cite{duch,
  newman_eigenvector, palla05}. The principal method for detecting communities
  is through modularity optimization, which boils down to clustering nodes based
  on the link densities within and between communities. The link densities
  should be high within communities and low between communities.

  The ordinary\footnote{Keep in mind that we consider directed graphs. Therefore
  the sum of all degrees is $m$, not $2m$, which is reflected in some minor
  changes to the original definitions~\cite{girvan_newman_modularity}; see
  also~\cite{leicht2007}.} definition of modularity for directed
  graphs~\cite{leicht2007} is
  \begin{equation}
   Q(\{\sigma\}) = \frac{1}{m}\sum_{s} m_{ss} - [m_{ss}],
  \end{equation}
  where $m_{ss}$ is the actual number of links within a community $s$, $\sum
  A_{ij} \delta(\sigma_i, s)\delta(s, \sigma_j)$, and $[m_{ss}]$ is the expected
  number of such links, $\sum p_{ij} \delta(\sigma_i, s)\delta(s, \sigma_j)$,
  where $p_{ij}$ is some expected value in a random null model. The expected
  values are constrained by $\sum p_{ij} = m$, because the random null model
  should have the same number of links as the actual network. Taking degrees
  into account, a sensible expectation is $p_{ij} = k^{out}_ik^{in}_j/m$,
  which was used in the original
  definition~\cite{leicht2007}.\footnotemark[\value{footnote}] 

\begin{figure}
    \begin{center}
      \includegraphics[]{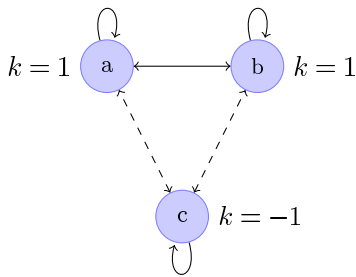}
      \caption{(Color online) Network illustrating the shortcoming of modularity
      when negative links are present. The dashed links have a weight of $-1$
      and the others have a weight of $1$. Using expected link values $p_{ij} =
      k^{out}_ik^{in}_j/m$, the expected values equal the actual values, $p_{ij}
      = A_{ij}$. Hence the modularity $Q = 0$ for all
      configurations.}\label{fig:problem_modularity} 
    \end{center}
  \end{figure}

  A straightforward generalization to weighted graphs is to set $A_{ij} =
  w_{ij}$ and to take the degree measure as the sum of the link weights. For
  graphs with negative weights, however, a problem arises, illustrated in
  Fig.~\ref{fig:problem_modularity}. The weighted degree of the three nodes $a$,
  $b$ and $c$ is $k_a = 1$, $k_b = 1$ and $k_c = -1$. The total weighted degree
  is $m = \sum w_{ij} = 1$. The expected values $p_{ij} = k^{out}_ik^{in}_j/m$
  equal the edge weights $w_{ij}$. Hence $A_{ij} - p_{ij} = 0$ for all links,
  and each possible community configuration results in a modularity $Q = 0$,
  while the appropriate configuration is clear from the figure: $a$ and $b$
  belong to the same community, and $c$ to another community ~\cite{gomez2009}.
  Some adaptations of modularity are therefore required to detect communities in
  networks with (also) negative links.

\section{Potts model and extension}

  To get rid of the shortcomings of frustration and modularity, we will attempt
  to modify the latter. After all, for many networks with only positive links,
  modularity has shown to uncover valid community structures in multiple
  areas~\cite{reichardt09}. We will extend the original definition of modularity
  to allow for negative links, by phrasing our challenge as a Potts
  model~\cite{wu1982}. This model will also show up a close connection between
  social balance and modularity.

  We start out by treating the positive and the negative links separately.
  Mimicking the approach taken by Reichardt and
  Bornholdt~\cite{reichardt_spinglass} we first define a Hamiltonian for the
  positive part, which represents the ``energy'' of a given configuration
  $\{\sigma\}$. We reward internal positive links by $a_{ij}$ and penalize
  absent internal positive links by $b_{ij}$, which leads to
  \begin{equation}
    \mathcal{H}^+(\{\sigma\}) = \sum_{ij} \left[ - a_{ij} A^+_{ij} + b_{ij}(1 - A^+_{ij}) \right]  \delta(\sigma_i, \sigma_j).
  \end{equation}

  Setting $a_{ij} = 1 - b_{ij}$ and $b_{ij} = \gamma^+ p^+_{ij}$, where
  $p^+_{ij}$ represents the expected (positive) link between $i$ and $j$, allows
  us to simplify the above equation to\footnote{Notice that if we have a
  weighted network, $a_{ij} = w_{ij} - b_{ij}$.}
  \begin{equation}
    \mathcal{H}^+(\{\sigma\}) = - \sum_{ij} (A^+_{ij} - \gamma^+ p^+_{ij})\delta(\sigma_i, \sigma_j).
  \end{equation}
  which is the Potts model analyzed by Reichardt and
  Bornholdt~\cite{reichardt_spinglass} if only positive links are present.  We
  define the negative part analogously, but now we penalize internal negative
  links and reward absent negative internal links,
  \begin{equation}
    \mathcal{H}^-(\{\sigma\}) =  \sum_{ij} (A^-_{ij} - \gamma^- p^-_{ij})\delta(\sigma_i, \sigma_j).
  \end{equation}
  The effect of the negative links on the energy of the entire configuration is
  opposite to the effect of the positive links. Combining the two Hamiltonians
  into one yields
  \begin{equation}
    \mathcal{H}(\{\sigma\}) = (1 - \lambda)\mathcal{H}^+(\{\sigma\}) + \lambda \mathcal{H}^-(\{\sigma\}),
  \end{equation}
  where $\lambda$ plays a similar role as in frustration, of balancing the
  effects of positive and negative links. As explained earlier, it makes sense
  to weigh the contributions of each part equally, thus $\lambda=1/2$. To
  illustrate, let us define a new matrix $B_{ij} = (1 - \lambda)A^+_{ij} -
  \lambda A^-_{ij}$, and construct the Hamiltonian for this altered network by
  setting $\lambda=1/2$.  Since the expected values for $B$ are $p'^+_{ij} = (1
  - \lambda)p^+_{ij}$ and $p'^-_{ij} = \lambda p^-_{ij}$, the Hamiltonian for
  $B$ is equivalent to the one for $A$, up to a multiplicative constant of $2$.
  So we may indeed set $\lambda=1/2$ and then simplify the above Hamiltonian
  (up to the multiplicative constant of $2$) to
  \begin{equation}
    \mathcal{H}(\{\sigma\}) = - \sum_{ij} \left[ A_{ij} - (\gamma^+ p^+_{ij} - \gamma^- p^-_{ij})\right] \delta(\sigma_i, \sigma_j),
    \label{equ:hamiltonian}
  \end{equation}
  which is the measure that we optimize to detect a community structure in
  networks with both positive and negative links. It can be easily seen that
  when the network is positive (and $\gamma^\pm=1$) we obtain
  \begin{equation}
    Q(\{\sigma\}) = -\frac{1}{m}\mathcal{H}(\{\sigma\}).
    \label{equ:modularity}
  \end{equation}
  So minimizing the Hamiltonian~(\ref{equ:hamiltonian}) is the same as
  maximizing modularity. In fact we just compare the original network to the
  appropriate negative link null model, which wasn't the case in the original
  modularity~\cite{girvan_newman_modularity} and in the Potts
  model~\cite{reichardt_spinglass}. 

  The simplest version of the expected values, $p^\pm_{ij}$, is obtained by just
  regarding the proportion of positive or of negative links in the network,
  $p^\pm_{ij} = m^\pm/n(n-1)$. If we want to take the degree distribution into
  account, then $p^\pm_{ij} = ~ \outpm{k}_i ~  \inpm{k}_j/m^\pm$. The modularity
  given in~\cite{gomez2009} also defines this negative link null model
  appropriately, and is a special case of ours.

  When $\gamma^+=\gamma^-=0$, the Hamiltonian~(\ref{equ:hamiltonian}) equals the
  frustration~(\ref{equ:frustration}) of the network, and if the network is also
  balanced and complete (no missing edges), minimizing the
  Hamiltonian~(\ref{equ:hamiltonian}) yields the same result as minimizing the
  frustration~(\ref{equ:frustration}). This can be pointed out by defining the
  probabilities by $p^\pm_{ij} = m^\pm/n(n-1)$, and by allowing the complete and
  balanced network to consist only of link values $A_{ij} \in \{-1, 1\}$. Then,
  as long as $\gamma^+m^+ - \gamma^-m^- < n(n-1)$, the coupling $A_{ij} -
  (\gamma^+ p^+_{ij} - \gamma^- p^-_{ij})$ between each positively associated
  pair of nodes is positive. Hence, the configuration produced by minimizing the
  Hamiltonian is the same as when minimizing the frustration.

\section{The ground state}

  Finding the actual minimum of the Hamiltonian---the so called ground
  state---is NP hard~\cite{np_complete}, and therefore only heuristic methods
  can be applied.  Our modularity~(\ref{equ:modularity}) can be easily
  integrated with existing algorithms for modularity optimization, such as
  eigenvector \cite{newman_eigenvector}, extremal optimization~\cite{duch}, fast
  unfolding~\cite{blondel_fast_unfolding}, or simulated
  annealing~\cite{guimera2005,reichardt_spinglass}. We opted for simulated
  annealing~\cite{simulated_annealing} to minimize the
  Hamiltonian~(\ref{equ:hamiltonian}) because it performs well in standard
  performance tests~\cite{comparing, cartography, reichardt_spinglass,
  benchmark}, although it's not the fastest
  algorithm~\cite{newman06,blondel_fast_unfolding}.  Here we will give a short
  overview of how to adapt the simulated annealing
  approach~\cite{reichardt_spinglass} to incorporate negative links.

  First, it's convenient to define the adhesion between community $r$ and $s$,
  similar to~\cite{reichardt_spinglass}, $$a_{rs} = (m^+_{rs} - m^-_{rs}) -
  ([m^+_{rs}] - [m^-_{rs}]),$$ where $m^\pm_{rs} = \sum
  A^\pm_{ij}\delta(\sigma_i,r)\delta(\sigma_j,s)$ is the actual number of arcs
  from $r$ to $s$ and $[m^\pm_{rs}] = \sum \gamma^\pm p^\pm_{ij}\delta(\sigma_i,
  r)\delta(\sigma_j,s)$ is the expected number of arcs from $r$ to $s$.
  Hamiltonian~(\ref{equ:hamiltonian}) can be rewritten accordingly,
  \begin{equation}
    \begin{array}{rccl}
      \mathcal{H}(\{\sigma\}) & = & & \displaystyle \sum_{s} a_{ss} \\
      & = & - & \displaystyle \sum_{r \neq s} a_{rs},
    \end{array}
    \label{equ:hamiltonian_adhesion}
  \end{equation}
  where $r$ and $s$ are communities in $\{\sigma\}$.

  In order to minimize the Hamiltonian, we consider the effect of moving a
  single node from one community to another, expressed in terms of adhesion.
  More specifically, moving node $v$ from $r$ to $s$ results in the following
  change,
  \begin{equation}
  \Delta \mathcal{H}(\sigma_v\!\!:\!r \to s) = (a_{vr} + a_{rv}) - (a_{vs} + a_{sv}),
  \label{equ:change_hamiltonian}
  \end{equation}
  where $a_{vr}$ is the adhesion between node $v$ and its complement in
  community $r$.  Let us write the mutual adhesion of a node $v$ and a community
  $r$ as $\alpha_v(r) = a_{vr} + a_{rv}$.  If, for the sake of argument, the
  mutual adhesion of $v$ and $s$ is larger than the mutual adhesion of $v$ and
  $r$, $\Delta \mathcal{H}(\sigma_v\!\!:\! r \to s) = \alpha_v(r) - \alpha_v(s)$
  decreases the Hamiltonian.  In other words, $v$ has more positive, or less
  negative, links than expected to $s$ than to $r$, and moving $v$ to $s$ would
  improve the configuration. 

  To each move we can assign a probability~\cite{simulated_annealing}
  $$\Pr(\sigma_v\!\!:\!r \to s) = \frac{\exp (\beta \Delta
  \mathcal{H}(\sigma_v\!\!:\! r \to s))}{\sum_{i} \exp (\beta \Delta
  \mathcal{H}(\sigma_v\!\!:\! r \to i))},$$ where $T$ is the temperature and
  $\beta = 1/T$. By slowly decreasing the temperature, the probability of moving
  to another state approaches the maximum possible, thereby forcing the system
  into its minimum energy, i.e. the ground state. Notice that in principle the
  probabilities are dependent on the total energy $\mathcal{H}(\sigma_v: r \to
  s)$, but since $\mathcal{H}(\sigma_v: r \to s) = \mathcal{H}(\{\sigma\}) +
  \Delta \mathcal{H}(\sigma_v\!\!:\! r \to s)$ we can simplify to the equation
  stated above.

  The algorithm iterates randomly over the nodes a number of times, after which
  the temperature is decreased to a lower temperature $T'$ stepwise, and usually
  (although not necessarily) $T'=0.99T$. The iterations and the lowering of the
  temperature are continued until there are no further (significant)
  improvements.  Any further changes would result in a higher energy, which we
  do not want, hence the resulting configuration of minimum energy is our
  solution $\{\sigma\}$.  Herein, for any set of nodes $u$, its mutual adhesion
  to its own community $s$ is stronger than to any other community $r$,
  $\alpha_u(s) \geq \alpha_u(r),$ which is clear when one looks at
  Eq.~(\ref{equ:change_hamiltonian}). 

  Furthermore, the cohesion, or self-adhesion, $a_{ss}$ of a community is always
  positive, $a_{ss} \geq 0$, and the mutual adhesion between two communities is
  always negative, $a_{rs} + a_{sr} \leq 0$. If the cohesion were negative, we
  could then move a set of nodes to another community and thereby decrease the
  energy, which would contradict the fact that the system is in the ground
  state.

  In fact these last two inequalities can be rephrased, which yields some
  insight into the effect of the parameters $\gamma^\pm$. If we assume, for
  analytic purposes, that $p^\pm_{ij} = p^\pm = m^\pm/n(n-1)$, the expected
  values become $[m^\pm_{rs}] = p^\pm n_rn_s$ for $r\neq s$ and $[m^\pm_{ss}] =
  p^\pm n_s(n_s-1)$, where $n_s$ is the number of nodes in community $s$.
  Writing this out we arrive at
  \begin{eqnarray}
   \nonumber \frac{m^+_{ss} - m^-_{ss}}{n_s(n_s - 1)} & \geq & \frac{\gamma^+ m^+ - \gamma^-m^-}{n(n - 1)} \\ 
   & \geq & \frac{(m^+_{rs} - m^-_{rs}) + (m^+_{sr} - m^-_{sr})}{2n_rn_s},
  \end{eqnarray}
  wherein the middle term is a sort of global density. Hence by changing
  $\gamma^\pm$ we change the threshold for clustering nodes together versus
  keeping them apart. Either way, the density within a community is always
  higher than the global density of the network, while the density between
  communities is always lower than the global density. Increasing $\gamma^+$
  raises the threshold for nodes to be clustered, and will (generally) result in
  smaller communities detected, possibly embedded in larger and sparser
  communities. Increasing $\gamma^-$ has the opposite effect and lowers the
  threshold, and will (generally) result in a configuration of larger
  communities.

\section{Application}

\begin{figure*}
  \begin{center}
    \includegraphics[width=\textwidth]{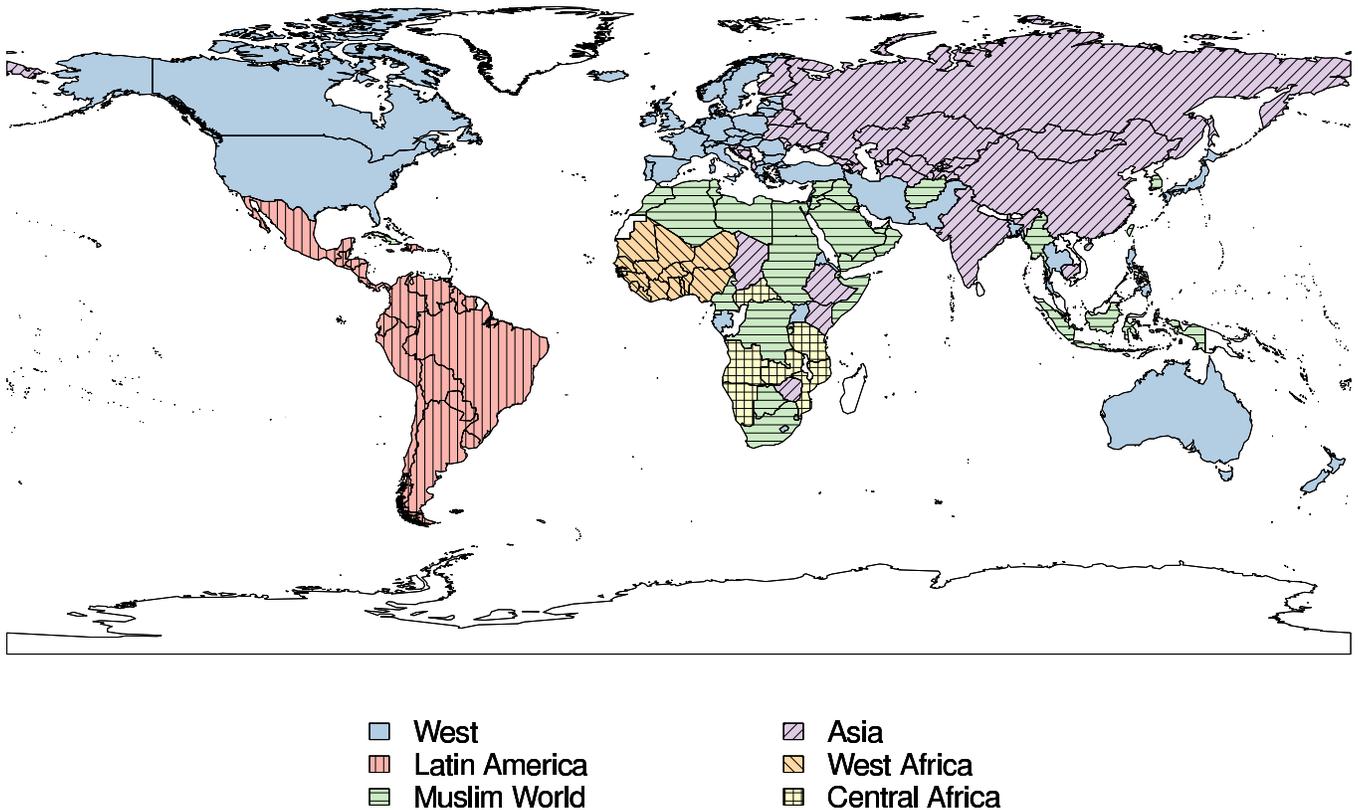}
    \caption{(Color online) Map of the communities in the conflict and alliance
    network found using the algorithm described in the text ($Q=0.561$,
    $\gamma^\pm = 1$).\label{fig:map}
    }
  \end{center}
\end{figure*}

  To show how our method can be applied to an empirical network, we analyze
  international relations taken from the Correlates of War~\cite{mid3,
  cow_alliance} data set over the period 1993--2001, where military alliances
  can be represented by positive links and disputes by negative links. The data
  set contains a wide variety of disputes,  for example border tensions between
  Colombia and Venezuela, the deployment of Chinese submarines to Japanese
  islands, and Turkish groups entering Iraqi territory.  Disputes were assigned
  hostility levels, from ``no militarized action'' to ``interstate war,'' and we
  chose the mean level of hostility between two countries over the given time
  interval as the weight of their negative link. The alliances we coded one of
  three values, for (1) entente, (2) non-aggression pact, or (3) defense pact.
  The disputes $w^-_{ij}$ and alliances $w^+_{ij}$ are both normalized to values
  in the interval $w^\pm_{ij} \in [0,1]$ which then bear equal weight in the
  overall link value $w_{ij} = w^ +_{ij} - w^-_{ij}$. The largest connected
  component consists of 161 nodes (countries) and 2517 links (conflicts and
  alliances).

  The result of the analysis ($Q=0.561$) is shown in Fig.~\ref{fig:map}.
  Countries of the same color (or pattern) belong to the same community, which
  in this context is more appropriately labeled a \emph{power bloc}.  How
  strongly a country belongs to its power bloc can be determined by the adhesion
  $\alpha_v(s)$ a node has to its community.  The power blocs can be identified
  as follows: (1) the West; (2) Latin America; (3) Muslim World; (4) Asia; (5)
  West Africa; and, (6) Central Africa.  If we detect communities by using only
  positive links, there is an agreement of about 64\% with the configuration in
  Fig.~\ref{fig:map}, while if using only negative links, there is an agreement
  of about 30\%. 

  Our result resembles the configuration depicted in Huntington's renowned book
  \emph{The Clash of Civilizations}~\cite{huntington}, with a few notable
  exceptions. The West African power bloc is an additional insight that is
  absent in Huntington's configuration.  A major difference with Huntington is
  that China itself does not constitute a separate bloc, nor does Japan or
  India. Some other noteworthy differences are Pakistan and Iran which are
  grouped with the West, while South Korea and South Africa are grouped with the
  Muslim World.

  If we run the algorithm with $\gamma^+=0.1$ and $\gamma^-=1$, North America
  merges with Latin America, while Europe becomes an independent community, and
  North Africa and the Middle East align with Russia and China. When setting
  $\gamma^+=1$ and $\gamma^-=2$, in contrast, former Soviet countries separate
  from Russia and form an independent community.  Using a range of values for
  $\gamma^\pm$, one can detect various layers in the community structure.

  Our configuration does not imply that conflicts take place between power blocs
  only, as 24\% of all conflicts actually take place \emph{within} blocs. For
  example, Georgia and Russia had serious conflicts, and DR of Congo and Rwanda
  had theirs, but each of these pairs is grouped together nevertheless. In these
  cases, the alliances overcame the conflicts in the grouping, confirming that a
  configuration of international relations is more than the sum of bilateral
  links.

  Our political analysis here is limited, since we wish to demonstrate the
  method rather than present a complete coverage of international alliances and
  disputes.  Other approaches that could be brought into play are the democratic
  peace theory~\cite{hensel, tocqueville}, which predicts few conflicts between
  democratic countries but fails to predict that in actuality, most conflicts
  occur between democratic and non-democratic countries; the realist
  school~\cite{kissinger}, which emphasizes geopolitical concerns; and, the
  trade-conflict theory~\cite{trade_conflict}, which argues that (strong) trade
  relations diminish the probability of a dispute, or lower its intensity.  In
  sum, although Huntington's configuration of civilizations was
  questioned~\cite{tucker,russett}, it seems to be fairly robust and with some
  marked exceptions is confirmed by our analysis.

\section{Conclusion}

  We have extended the existing Potts model by adapting the concept of
  modularity to detect communities in complex networks where both positive and
  negative links are present.  This approach solves a long-standing problem in
  the theory of social balance, namely the clustering of signed graphs.

  As a case in point, we have analyzed a social network of international
  disputes and alliances.  Other applications could be networks of references on
  the Web~\cite{flake} or in blogs~\cite{adamic}. If in these data positive and
  negative references are distinguished, our method makes possible to detect not
  only thematic clusters, but also positional clusters with internal agreement
  and external disagreement.
    
  For network data, the model's parameters ($\gamma^\pm$) can be used to find
  smaller (sub) communities, although there is currently no theoretical guidance
  to choose parameter values \cite{kumpala07}.  Even if there were such
  guidance, the modularity approach intrinsically aims at global rather than
  local optimization.  Our implementation is based on simulated
  annealing~\cite{reichardt_spinglass, simulated_annealing}, which performs
  quite well with standard tests, although for very large networks, faster
  algorithms will be necessary~\cite{blondel_fast_unfolding}.

  Whatever algorithms future researchers will use, or improvements of the
  concept of modularity they will develop, being able to detect communities in
  networks with both positive and negative links is important in numerous fields
  of science, and a stepping stone toward further analyses of complex networks.

\begin{acknowledgements} 
  The authors like to thank anonymous referees and Jean-Charles Delvenne for
  their valuable comments and advice. VT acknowledges support from a grant
  ``Actions de recherche concertées – Large Graphs and Networks'' of the
  ``Communaut\'e Fran\c caise de Belgique'' and from the Belgian Network DYSCO
  (Dynamical Systems, Control, and Optimization), funded by the Interuniversity
  Attraction Poles Programme, initiated by the Belgian State, Science Policy
  Office.  
\end{acknowledgements}

\end{document}